\documentclass[11pt]{article}

\usepackage[final]{acl}

\usepackage{times}
\usepackage{latexsym}
\usepackage{amsmath}
\usepackage[T1]{fontenc}

\usepackage[utf8]{inputenc}

\usepackage{microtype}

\usepackage{inconsolata}

\usepackage{graphicx}

\usepackage{booktabs}
\usepackage{multirow}
\usepackage{amssymb}
\usepackage{pifont}

\usepackage[capitalize]{cleveref}
\crefformat{section}{\S#2#1#3}
\crefformat{subsection}{\S#2#1#3}
\crefformat{subsubsection}{\S#2#1#3}

\definecolor{DarkGreen}{RGB}{80,134,84}

\usepackage{xspace}
\newcommand{\masp}{\texttt{UNIVID}\xspace}

\newcommand{\masplite}{\texttt{UNIVID}-Lite\xspace}

\usepackage{enumitem}
\usepackage[dvipsnames]{xcolor}
\usepackage{caption}
\usepackage{placeins}
\usepackage{bm}
\usepackage{booktabs}
\usepackage{graphicx}
\usepackage{multirow} 
\usepackage{array}  
\usepackage[linesnumbered,ruled,vlined]{algorithm2e}
\usepackage{amsmath} 
\usepackage{blindtext}
\usepackage{etoolbox,lipsum}
\usepackage[table]{xcolor}

\usepackage[many]{tcolorbox}
\usepackage{marginnote}
\tcbuselibrary{skins,breakable}
\newtcolorbox{story}[1][]{
  enhanced,
  breakable,
  pad at break=2mm,
  left=2mm,right=2mm,
  colback=white,colframe=black!50!black,
  drop fuzzy midday shadow=black!50!black, 
  width={\columnwidth}, 
  frame hidden, 
  segmentation hidden,
  fontupper=\ttfamily, 
  before=\par\vspace*{2mm},after=\par\bigskip,
  title=#1}%

\usepackage{colortbl} 
\definecolor{lightgreen}{RGB}{198,239,206} 
\makeatletter
\renewcommand{\paragraph}{%
  \@startsection{paragraph}{4}{\z@}{1ex}{-1em}{\normalfont\normalsize\bfseries}%
}

\newcommand{\hide}[1]{}

\title{\masp: Unified Vision-Language Model for Video Moderation}

\author{
  \textbf{Kejuan Yang\textsuperscript{*}},
  \textbf{Yizhuo Zhang\textsuperscript{*}},
  \textbf{Mingyuan Du},
  \textbf{Yue Zhang},
\\
  \textbf{Dixin Zheng},
  \textbf{Kaili Zhao},
  \textbf{Yang Xiao},
  \textbf{Hanzhong Liang},
  \textbf{Kenan Xiao}
\\
\\
  Bytedance
}

\makeatletter
\def\blfootnote{\xdef\@thefnmark{}\@footnotetext}
\makeatother

\begin{document}
\maketitle

\blfootnote{\textsuperscript{*}Equal contribution.}
\blfootnote{Correspondence: \href{kendra.kejuanyang@bytedance.com}{kendra.kejuanyang@bytedance.com}}

\begin{abstract}
Global-scale video moderation faces a dual challenge: the need for fine-grained multi-modal reasoning and the demand for interpretable outputs to support downstream enforcement.
Traditional moderation systems often rely on fragmented black-box classifiers that are difficult to maintain and lack transparency. 
In this paper, we present \masp, a
\textit{\textbf{UNI}fied \textbf{VI}sion-language model for video mo\textbf{D}eration}.
Unlike standard classification models, \masp generates policy-aware captions that serve as an interpretable intermediate representation, enabling human-verifiable decisions and multi-task reusability. 
While existing open-source and commercial VLMs often suffer from safety-guardrail refusals and lack fine-grained policy alignment, we develop a specialized training data recipe that combines expert human-refined labels with synthetic data to align the model with our safety guidelines.
By integrating UNIVID as the core captioner, we design a novel end-to-end video moderation system that reduces violation leakage by 42.7\% and overkill rate by 37.0\% relatively. 
Meanwhile, by replacing over 1,000 policy-specific models with a single \masp backbone, we recycled extensive computation resources while reducing engineering maintenance overhead.
To our knowledge, this is one of the first reports of a high-efficiency captioning VLM successfully supporting industrial-scale moderation and cross-functional business.
\end{abstract}

\section{Introduction}

The rapid growth of short-video platforms such as Reels and YouTube Shorts necessitates accurate and operationally efficient moderation systems~\cite{wang2025filter, liang2025embedding}.
Prior systems rely on thousands of specialized, end-to-end classification models ~\cite{shi2024cpfd}, each targeting a specific policy (e.g., violence, regulated activities). However, this approach faces three bottlenecks: (1) Lack of Interpretability, as black-box scores offer little to no rationales for human auditors ~\cite{levi2025ai, bao2025vmoda}; (2) Tedious Maintenance, where policy updates require retraining thousands of individual models ~\cite{liang2025embedding}; and (3) Resource Inefficiency, as these models lack the semantic flexibility to support cross-functional business such as advertisement and recommendation safety.

Vision-Language Models (VLMs) offer a promising alternative by transforming video into natural language descriptions, i.e., video captions. \textbf{Why using a Captioning VLM for moderation?} 
First, captions act as a unified, human-readable bridge that provides explicit evidence for policy violations ~\cite{huang2025llavashieldsafeguardingmultimodalmultiturn}. 
Second, a single VLM can replace thousands of specific classification models ~\cite{Wang2026FromNM}, simplifying our system infrastructure. 
Finally, these policy-aware captions can be cached and reused by downstream tasks, creating a multi-purpose content understanding feature ~\cite{wang2025vips}.

Despite their potential, existing open-source ~\cite{liu2023visual, liu2024llava, li2024llavaov} and commercial VLMs ~\cite{comanici2025gemini, gpt4o} often fall short of industrial moderation requirements. They frequently refuse to describe sensitive or violative content ~\cite{bao2025vmoda, lee2025vision} due to internal safety triggers. Furthermore, since these models are not grounded using our internal platform policies, they often fail to capture the precise enforcement boundaries, leading to inaccurate descriptions. 
Finally, the scale of commercial VLMs makes real-time inference for full-traffic moderation economically impractical.


To address these limitations, we develop \masp, which adopts the same architectural design as LLaVA-OV~\cite{li2024llavaov} while employing a task-specific training recipe for moderation.
We use a hybrid training strategy that combines expert annotations with high-quality synthetic data, aligning the model with our detailed moderation policies.
On top of \masp, we construct a multi-stage moderation system: 
a high-throughput Risk Filter leveraging \masp embeddings for early screening; 
a Moderation Actor deploying two finetuned variants of \masp to support moderation decisions; 
and a Trend Governance module that reuses cached \masp captions to detect emerging risks.

Our \masp-centric moderation system has been fully integrated into our platform, yielding the following contributions:

\begin{itemize}[leftmargin=10pt, labelsep=5pt, itemsep=0pt]
    \item \textbf{Unified System}: We propose the industry deployment of a unified video moderation infrastructure built on \masp, significantly reducing engineering overhead and simplifying end-to-end troubleshooting.
    \item \textbf{Data \& Evaluation Pipeline}: We design a human-in-the-loop training recipe to ensure factual and policy alignment.  Furthermore, we introduce \textit{CapBench} for evaluation, which decomposes captions into atomic events to evaluate violation recall across key safety domains.
    \item \textbf{Platform Safety Governance}: By integrating UNIVID as the core captioner, our new moderation system reduces violation leakage by 42.7\% and overkill rate by 37.0\% relatively. Beyond that, \masp also achieves 81\% matching accuracy on the beta simulation of Brand \& Ads applications.
\end{itemize}

\section{Method}

\subsection{Model Architecture}

\begin{figure}[t] 
  \centering
  \includegraphics[width=0.8\linewidth]{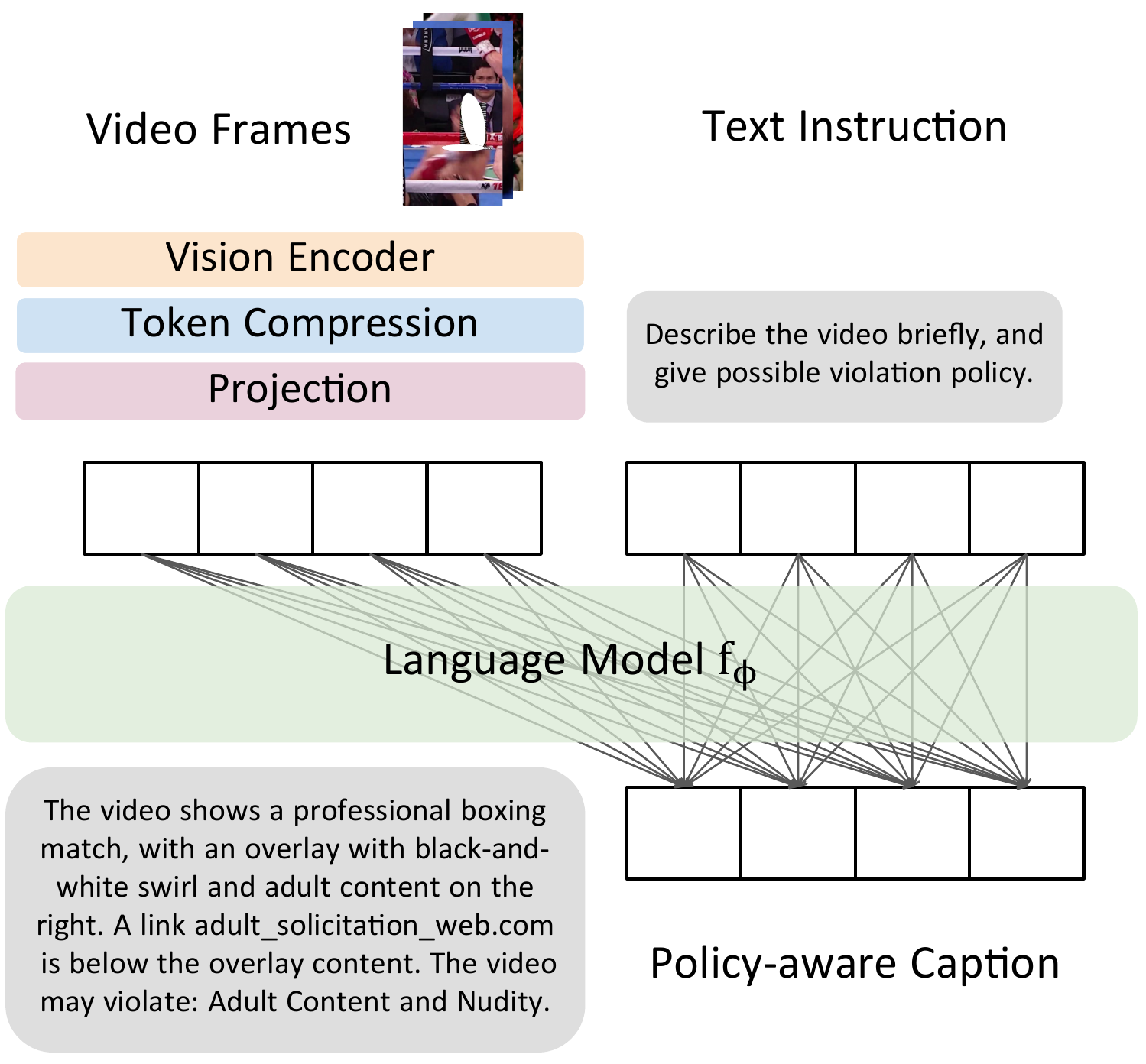}
  \caption{Model architecture of \masp following LLaVA-OneVision~\cite{li2024llavaov}. We construct in-house data recipe focusing on safety violation content.}
  \vspace{-10pt}
  \label{fig:model_arch}
\end{figure}

\begin{table*}[t]
\centering
\caption{Overview of \masp training stages.}
\resizebox{0.85\textwidth}{!}{%
\begin{tabular}{@{}lllc@{}}
\toprule
Stage               & Data Source                          & Type                                       & \# Samples (M)  \\ \midrule
PT & LLaVA pretrain, our internal data & Single sentence caption & 1.6 \\ \midrule
\multirow{2}{*}{FT} & Synthetic and human-refined caption & Caption                                    & 3.2 \\
                    & Synthetic VQA                     & General VQA (e.g., summary, topic, keywords, etc.) & 2.0  \\ \midrule
CFT                 & Hybrid caption                  & Caption                                    & 0.1  \\ \bottomrule
\end{tabular}
}
\label{tab:train_stage_and_data}
\vspace{-10pt}
\end{table*}

Our \masp model is built upon the LLaVA family of multimodal large language models \cite{liu2023visual, liu2024improved, liu2024llava, li2024llavaov}, with a particular focus on the LLaVA-OneVision \cite{li2024llavaov} architecture, which has demonstrated strong scalability and flexibility for video content understanding. We choose Mistral-v0.3-7B \cite{Jiang2023Mistral7} as the LLM $f_{\phi}$, as its permissive license ensures production compliance and its parameter scale is optimized for full-traffic deployment. We follow LLaVA-OneVision \cite{li2024llavaov} for design choices on components shown in Figure \ref{fig:model_arch}. 

To be more specific, for a video input $V$, and the associated caption $C$, we first extract frames uniformly from the video, providing a series of video frames $X_{1...N}$, where $N$ is the maximum supported frames for our model. After appropriate pre-processing, the image is fed into vision encoder $g$, token compression module $TC$, and projection $W$. This process eventually convert visual information into tokens $H_v$, which is then concatenated with encoded text instructions $H_t$. For a caption $C$ with length $L$ we aim to maximize the following:

\vspace{-15pt}
\begin{equation*}
    P(C \mid T_{\text{in}}, V_{\text{in}}) = \prod_{i=1}^{L} P(C_{i} \mid H_{\text{t}}, H_{\text{v}}, C_{{<i}})
\end{equation*}

\subsection{Data Recipe}



Constructing a large-scale training set for video moderation poses several challenges.
Firstly, collecting real-world violative video is difficult and legally sensitive. Therefore, no open-source dataset is available. 
Moreover, we cannot completely rely on proprietary or open-source VLMs, which are safety-aligned and frequently refuse to generate sensitive labels, e.g., CSAM\footnote{CSAM: Child Sexual Abuse Material}. 



To address these limitations, we develop an in-house dataset tailored to the specific content styles and moderation requirements of our platform (see Table~\ref{tab:train_stage_and_data}). Following the standard LLaVA instruction-tuning recipe~\cite{liu2024llava}, we leverage GPT-4o~\cite{gpt4o} to generate one-paragraph captions and VQA pairs\footnote{GPT-4o was the state-of-the-art VLM in multimodal understanding when we designed the system.}. The VQA pairs encompass both general visual understanding and moderation-specific queries. Examples of training data are shown in Appendix \ref{app:training-data}.


To optimize the trade-off between annotation costs and data quality, we internally train annotators to perform human-in-the-loop refinement of the raw GPT-4o outputs. The refinement includes two dimensions:
(1) \textbf{Factual Correction}: Rectifying hallucinations and adding missing details regarding subjects, objects, actions, backgrounds, and OCR.
(2) \textbf{Policy Grounding}: Ensuring that violation rationales are grounded in our internal safety policy playbook. For violative content, annotators are required to map the video to the corresponding pre-defined policy title.


Furthermore, to better capture edge cases in harmful trends, we use proprietary VLMs to normalize and enrich the violation rationale description. This step specifically targets multilingual OCR and adversarial visual hacking techniques (e.g., split-screen effect).

\begin{table*}[htbp]
\centering
\caption{Comparison of violative video captioning benchmarks.}
\renewcommand{\arraystretch}{1.15}
\resizebox{0.9\textwidth}{!}{%
\begin{tabular}{llcccc}
\hline
Benchmark & \# Total (Vio.) & Human-verified & Global Source & Uni-scoring & Credibility \\ \hline
Dream-1k~\cite{dream1k}                            
& 1000 (0) 
& \checkmark 
& \ding{55} 
& \ding{55} 
& \checkmark \\

KuaiMod~\cite{kuaimod} 
& 1000 (422)             
& \checkmark 
& \ding{55} 
& \ding{55} 
& \ding{55} \\

\textbf{CapBench}~(Ours)                     
& 17210 (11476)      
& \checkmark 
& \checkmark 
& \checkmark 
& \checkmark \\ \hline
\end{tabular}
}%
\label{tab:benchmarks}
\vspace{-5pt}
\end{table*}

\subsection{Training details}
Following the LLaVA setup, our model training paradigm includes three stages: 
(1) Pretraining for modality alignment, 
(2) Instruction-tuning on caption and VQA data,
(3) Continue finetuning on high-quality caption data to further enhance the model's generation ability (see Table~\ref{tab:train_stage_and_data}).

In the pretraining stage, we freeze the whole model except the MLP projector to align the vision and text modality. In the next two finetuning stages, we train both the projector and LLM decoder.
The model training procedure takes 120 hours on 32 H100 GPUs.
We also develop \masp-1B as a lightweight variant by replacing the decoder with our proprietary LLM.

\subsection{Video Moderation System}

\masp functions as the core of our moderation system across following three stages. 


\hide{
Traditional architectures often rely on tightly coupled components, leading to costly and inflexible retraining cycles. In contrast, our pipeline leverages \masp as a centralized foundation model to decouple content understanding from policy enforcement. As illustrated in Figure~\ref{fig:main_pipeline}, \masp serves as a shared feature provider and can also be fine-tuned as a ranker for final enforcement. To balance inference efficiency with moderation agility, we organize the system into three stages:
}

\paragraph{Risk Filter}
Our risk filter module performs early-stage screening over full video traffic to identify potential policy violations while maintaining low latency and high throughput. 
We first generate video captions using \masp, augmented by auxiliary signals including OCR, titles, and risk embeddings predicted by lightweight policy-specific modules. 
Our fusion network integrates these multimodal inputs into a shared embedding, which is connected to multiple MLP policy heads. Each policy head applies a calibrated decision threshold to identify high-risk videos and route them to downstream stages.
    
\hide{
Serving as filter, this stage aggrates multiple text signals—including OCR text, titles, auxiliary risk signals, and MASP-generated video captions through a transformer-based fusion network cascaded with classifiers to estimate risk scores for different issue categories to each video. Low-risk content is filtered out, while ambiguous or high-risk cases are forwarded to the ranking module. Crucially, the filter sub-system only rely on MASP-generated captions and other text-based features; because MASP produces captions in a consistent style, updates to \masp do not disrupt the filter’s behavior.
}
    
\paragraph{Moderation Actor}
Our earlier moderation system suffered from precision dilution and high maintenance overhead. 
This is due to the reliance on multiple independent components (e.g., neural scoring models, vector-based retrievers, and heuristic rules) operating in parallel. 
Moreover, such a fragmented system complicates the threshold calibration and makes end-to-end troubleshooting tedious. To address these limitations, we transition to a recall-and-rank architecture composed of two models: UNIVID-Lite for primary enforcement and UNIVID-RAG for leakage mitigation.



\subparagraph{UNIVID-Lite}
 UNIVID-Lite serves as a unified actor model, providing a single decision layer for the moderation pipeline. It is finetuned on 1 million in-house videos based on the UNIVID-1B backbone. The training data consists of human-annotated moderation examples balanced at a positive-to-negative ratio of 1:5 to reflect the natural distribution of production traffic. Each sample includes video frames together with auxiliary textual signals which are concatenated directly into the instruction prompt (see Appendix D.2). UNIVID-Lite is trained with an autoregressive generation objective: rather than using a separate classification head, the model is prompted to produce a structured natural language output indicating the moderation decision (Approve or Violation) along with the corresponding violated policy. This formulation allows the model to inherit the general reasoning capabilities of the pretrained UNIVID backbone while specializing for binary enforcement.




\subparagraph{UNIVID-RAG}
 To mitigate leakage — violations missed by the primary actor — we introduce UNIVID-RAG, which augments the moderation decision with retrieval from a Violation Knowledge Base (VKB). The VKB contains approximately 100,000 structured violative events derived from past cases labeled by human moderators. When new violations are identified, Gemini-2.5-Pro converts the raw annotations into structured violative events capturing the policy title and violation rationale, which are then added to the VKB. The knowledge base is updated continuously to reflect the latest labeling.

At inference time, UNIVID-RAG retrieves the top-3 most semantically similar violative events from the VKB using cosine similarity over UNIVID caption embeddings. The retrieved events are inserted directly into the prompt as in-context examples, providing the model with concrete prior cases to reason against (see Appendix D.3). This retrieval-augmented approach is specifically designed to improve coverage of hard or low-frequency violations that the primary model may miss, at the cost of a moderately higher violation rate due to increased sensitivity.

\hide {
The enforcement stage uses MASP-Lite, a distilled 1B-parameter variant of the foundation model that serves as an MLLM-based ranker. This stage focuses on the remaining high-risk samples and applies fine-grained reasoning to identify policy violations. To address particularly challenging cases, we additionally build a knowledge base that is continuously updated by human labelers. This knowledge base stores historical hard cases along with human annotations, and we use Gemini to generate structured reasoning for these cases. During enforcement, the system retrieves similar examples and leverages the associated knowledge to support more accurate and consistent final decisions.
}

\paragraph{Trend Governance}


Short-form video platforms are increasingly targeted by emerging trends, such as dangerous ``hot water pouring challenges''. 
To identify these real-time issues, we implement a lightweight adaptation strategy. 
By reusing cached fusion embeddings from the backbone, we train an MLP trend head to recognize these shared semantic videos via few-shot samples (< 50). 
This approach allows our moderation system to remain agile, capturing short-lived or rapidly evolving threats.

\begin{table*}[tb]
\caption{Evaluation results on CapBench.}
\renewcommand{\arraystretch}{1.13}
\resizebox{\textwidth}{!}{%
\begin{tabular}{@{}lcccccccccc@{}}
\toprule
                & \multicolumn{7}{c}{\textbf{Violative Set}}                                      & \multicolumn{3}{c}{\textbf{Healthy Set}} \\ 
 & Violence & Sex Abuse & Mental Health & Regulated Act & Integrity & Vio Rec. & Non-vio Rec. & Rec. & Prec. & F1 \\ \cmidrule(l){2-8} \cmidrule(l){9-11}

\rowcolor{gray!20} \textit{Proprietary Models} & & & & & & & & & & \\ 
\rowcolor{gray!20} GPT-4.1         & 45.9 & 17.4 & 32.3 & 42.5 & 57.6 & 36.1 & 32.8 & 31.0 & 65.5 & 37.4 \\
\rowcolor{gray!20} Gemini-2.5-Pro & 63.8 & 44.3 & 55.6 & 57.6 & 67.5 & 55.1 & 42.5 & 44.9 & 95.2 & 57.9 \\ \midrule

LLaVA-OV 8B     & 17.8 & 6.9  & 12.9 & 15.7 & 14.2 & 13.0 & 12.0 & 12.7 & \textbf{86.3} & 19.3 \\
\textbf{\masp-7B}  & \textbf{56.3} & \textbf{51.3} & \textbf{50.2} & \textbf{57.7} & \textbf{50.1} & \textbf{54.3} & \textbf{32.4} & \textbf{28.9} & 82.3 & \textbf{39.1} \\
\masp-1B & 53.6 & 49.1 & 49.9 & 55.3 & 47.7 & 52.1 & 30.6 & 27.4 & 82.8 & 37.5 \\
\midrule
\textit{Ablation Study} \\
w/o Hybrid Data               & 37.9 & 35.5 & 33.5 & 41.1 & 30.4 & 37.5 & 18.4 & 15.8 & 81.8 & 23.1 \\
w/o Human Data        & 29.1 & 22.9 & 18.9 & 29.4 & 23.9 & 26.1 & 15.8 & 15.8 & 82.2 & 23.0 \\
\bottomrule
\end{tabular}
}
\vspace{-10pt}
\label{tab:capbench-main}
\end{table*}

\subsection{Evaluation}

Existing video safety benchmarks are often limited to surveillance contexts~\cite{sultani2018real, hassner2012violent} or synthetic video sources~\cite{videosafetybench}, failing to capture the broad spectrum of user-generated content. 
While the recent KuaiMod~\cite{kuaimod} benchmark addresses some of these gaps, it is localized to Chinese content and employs compliance-driven masking of key visual components, such as human faces.
Such anonymization brings a distribution shift that deviates from the high-fidelity and unmasked real-world video posts.

To evaluate multimodal understanding across both general and harmful contexts, we design our in-house benchmark \textbf{CapBench}, which assesses VLMs on two dimensions: descriptive accuracy and violation recall.
Inspired by the prior Dream-1k~\cite{dream1k} benchmark, we employ a fine-grained metric that decomposes video captions into atomic ``evidence.''
CapBench includes two subsets: \textbf{violative} and \textbf{healthy} sets. 
The violative subset is further divided into five domains: \textit{Violence, Sexual Abuse, Mental Health, Regulated Activity,} and \textit{Integrity}. Details are shown in Appendix~\ref{app:capbench-details}.

For each video, we establish a human-refined ground-truth caption, which is then segmented by Gemini-2.5-Pro into verifiable events. 
We then prompt Gemini as judge to compute two symmetric metrics based on these segments: 
(1) Recall: Measures the coverage of ground-truth events within the predicted caption.
(2) Precision: Measures the rate of hallucination or the proportion of predicted events supported by the ground truth.
For violative samples, events serving as violation rationales are annotated with violation labels (e.g., ``\textit{The three men ride together on a single red motorcycle.}''), enabling separate evaluation of violative and non-violative recall.
Compared to previous work in Table~\ref{tab:benchmarks}, our benchmark offers clearer insights into caption credibility and provides a unified scoring framework applicable to both violative and healthy video captioning.


\section{Results}

This section introduces our offline evaluation results and the simulated online impact of integrating \masp into our video moderation system.

\subsection{Offline Experiments}
We validate the captioning and violation recognition ability of \masp on our internal CapBench. 
\paragraph{Experiment Settings}
 We evaluate the leading commercial VLMs, specifically GPT-4o~\cite{gpt4o} and Gemini-2.5-Pro~\cite{comanici2025gemini} to explore the upper bound performace. Additionally, we benchmark LLaVA-OneVision-1.5 8B~\cite{LLaVA-OneVision-1.5}, which shares a similar architecture with \masp but has not been finetuned on our in-house moderation dataset.
 The model generation outputs are constrained to 150 words to align with our online deployment, as shown in Appendix~\ref{sec:appendix-prompts}.

\paragraph{Analysis}
As demonstrated in Table~\ref{tab:capbench-main}, \masp-7B substantially outperforms LLaVA-OV 8B, despite their similar architectures. 
Notably, \masp-7B also achieves higher violation recall across all domains compared to GPT-4.1~\cite{gpt4o}. 
This gap is partially due to the safety guardrails of proprietary models, which refuse generation in certain cases (12.9\%), yielding empty captions.

We further conduct ablation studies upon training data recipe.
We find that removing hybrid training data leads to consistent violation recall degradation across all domains (-16.8\% average). 
We observe a more pronounced decline when training with human-annotated data only, where violation recall further decreases to 26.1\%, indicating limited generalization under narrowly distributed supervision using GPT-4o.
Note that other ablation results on model architecture are presented in Appendix~\ref{app:masplite-casestudy}.


\begin{table}[tb]
\centering
\caption{Comparison of VLM deployment cost (per 1M videos). Note that we are unable to deploy LLaVA-OV due to compliance constraints.}
\vspace{-5pt}
\renewcommand{\arraystretch}{1.15}
\resizebox{\columnwidth}{!}{%
\begin{tabular}{lcccc}
\toprule
Metric 
& GPT-4.1 
& Gemini-2.5-Pro 
& LLaVA-OV 
& \masp \\ \midrule
\textbf{Vio Rec.} 
& 36.1 
& 55.1 
& 13.0 
& 54.3 \\

\textbf{Cost} (\$)
& 4830 
& 3444 
& N/A 
& 180 \\ 
\bottomrule
\end{tabular}
}%
\label{tab:cost-comparison}
\vspace{-15pt}
\end{table}

\subsection{Sandbox Moderation Experiments}

\masp has delivered substantial improvements in moderation accuracy on production-scale traffic. Based on our simulated evaluation results, our new system achieved a 42.7\% relative reduction in view-weighted violation leakage, falling from 0.255\% to 0.146\%. Simultaneously, the overkill rate decreased from 35.4\% to 22.3\%. 
Meanwhile, \masp replaces over 1000 policy-specific models with a single multimodal backbone, with 1900 A30 GPUs recycled. Consequently, our system reduces engineering overhead and simplifies long-term maintenance.

\paragraph{Deployment}
To support full video traffic, we deploy \masp model with FP8 quantization on H100 GPUs, reaching 5.7 QPS per device. 
As shown in Table~\ref{tab:cost-comparison}, our production inference cost is approximately \$180 per 1M videos, achieving a 15$\times$ reduction compared to commercial VLMs (>~\$3,000).

\subsection{Downstream Modules}

This section presents our simulated evaluation results for downstream systems that integrate \masp, including our video moderation main system and cross-functional applications.

\paragraph{Risk Filter} 
As shown in Table~\ref{tab:alice-e2e-eval}, integrating \masp embeddings consistently improves violation recall (especially +23.9\% on high-view leakage), highlighting enhanced sensitivity to rare but high-risk content.

After launching the Risk Filter with \masp embedding, our main moderation system observes an 11.5\% increase in violation hit volume. Meanwhile, the moderation precision improves substantially, with violation precision from 35.9\% to 75.6\%, demonstrating better coverage and decision quality in live traffic.

\paragraph{Moderation Actor}

Our simulated moderation ablation in Table~\ref{tab:ranking-masp-lite} demonstrates the effectiveness of \masp-Lite as a unified action layer in live moderation traffic. 
Compared to the recall-only pipeline, \masp-Lite reduces the violation rate (1.38\% → 1.34\%) while improving precision (+9.4\%) and leakage recall (+11.8\%), indicating more accurate enforcement with fewer unnecessary actions.

Introducing retrieval augmentation further increases leakage recall to 53.6\%, improving coverage of hard or low-frequency violations. 
This gain comes at the cost of slightly reduced precision and a slightly higher violation rate, reflecting the trade-off between maximizing recall and enforcing conservatively.
We further present our case study in Appendix~\ref{app:masplite-casestudy}.
Overall, \masp-Lite offers strong precision-oriented actions, with RAG serving as a complement when higher recall is prioritized.

\begin{figure}[t] 
  \centering
  \includegraphics[width=\columnwidth]{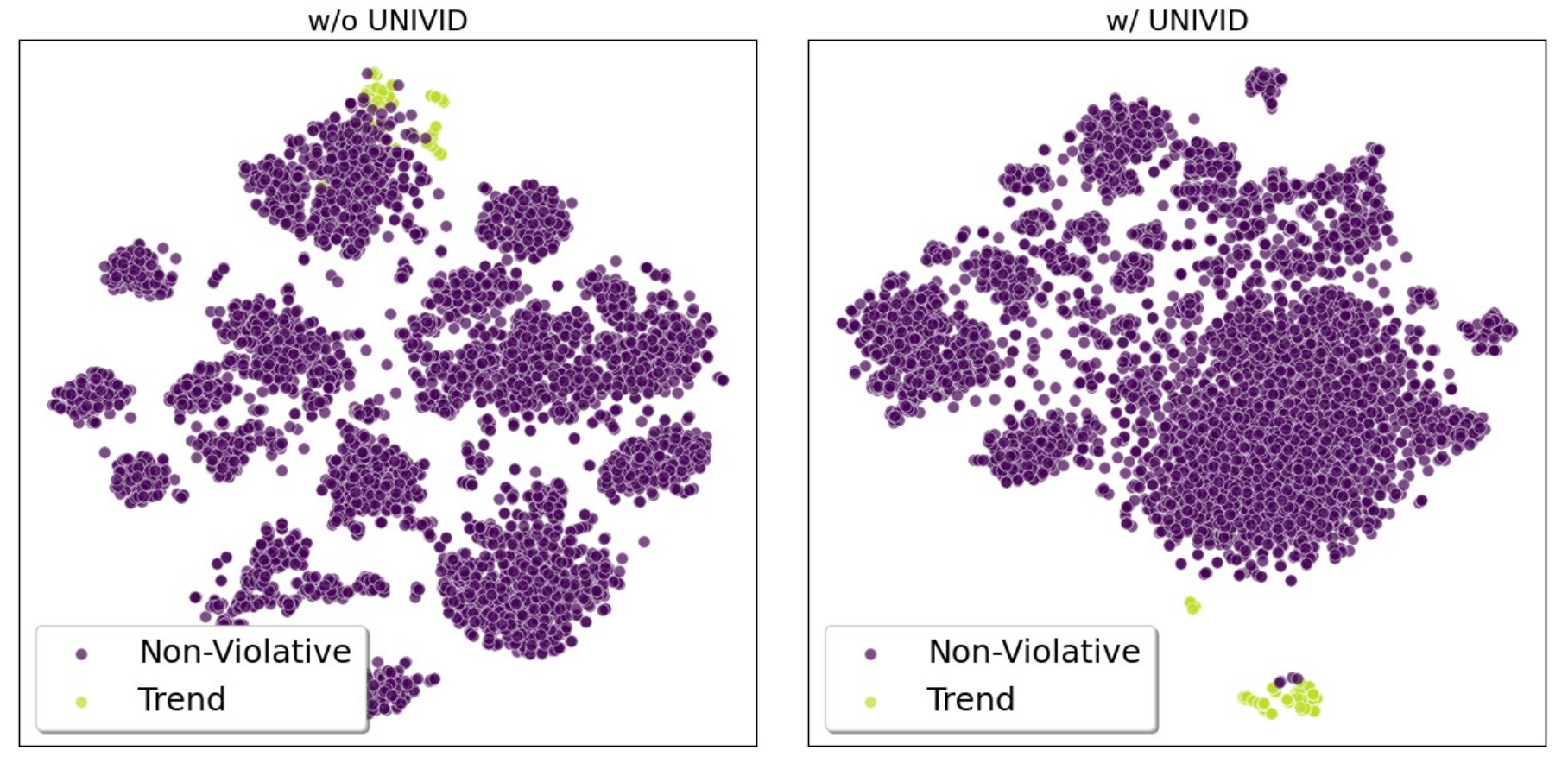}
  \caption{Visualization of trend detector embeddings.}
  \vspace{-2pt}
  \label{fig:trend-vis}
\end{figure}

\paragraph{Trend Governance}

We report the simulated moderation results of few-shot learning in Table~\ref{tab:alice-e2e-eval}. 
On the school dangerous behavior trend, incorporating \masp embeddings substantially improves moderation performance, boosting violation recall from 56.7\% to 86.7\%.
The embedding visualization in Figure~\ref{fig:trend-vis} further shows that \masp embeddings produce clearer separation between trend and non-violative cases.
This suggests that \masp provides semantically rich representations that generalize in low-data regimes, enabling rapid adaptation to emerging trends without degrading precision.

\paragraph{Cross-functional Business}

Beyond moderation, \masp also generates fine-grained video keywords that are consumed by the advertising ranking system as semantic targeting signals, enabling brand-safe matching and custom-lineup construction. This integration achieved 81\% accuracy in the Brand \& Ads applications in our simulation tests.

\begin{table}[tb]
\centering
\caption{Simulated online ablation of \masp embedding on the risk filter and few-shot trend detector. Scores are reported at 65\% precision threshold.}
\vspace{-5pt}
\renewcommand{\arraystretch}{1.12}
\resizebox{0.9\columnwidth}{!}{%
\begin{tabular}{lccc}
\toprule
Embeddings 
& Vio Rec. 
& Leak. Rec. 
& Trend Rec. \\ \midrule
w/o \masp 
& 72.3 
& 33.3 
& 56.7 \\

w/ \masp 
& \textbf{78.2}
& \textbf{59.8}
& \textbf{86.7} \\ 
\bottomrule
\end{tabular}
}%
\label{tab:alice-e2e-eval}
\vspace{-5pt}
\end{table}

\section{Related Work}
\subsection{Video Moderation Systems}

Video moderation is crucial for protecting users from detrimental content and maintaining a healthy platform, especially for minors~\cite{Gorwa2020Algorithmic,Gongane2022survey,Udupa2023Ethical,Lai2022Human-AI}.
Plenty of efforts have been made to construct a video moderation ecosystem.
Early work uses traditional ML methods on handcrafted features for hate speech, toxicity, or fake news detection~\cite{Gongane2022survey,Naseeb2025Machine}, typically in binary or multi-class setups. Similar patterns appear in visual moderation, where porn or anomaly detection in images/videos uses engineered visual features and classical classifiers~\cite{Wang2023Validating, Yousaf2022A}.


\begin{table}[tb]
\centering
\caption{Simulated online ablation results of \masp-Lite and \masp-RAG on the moderation actor module.}
\vspace{-0.5em}
\renewcommand{\arraystretch}{1.15}
\resizebox{0.48\textwidth}{!}{%
\begin{tabular}{lccc}
\toprule
Actor 
& Vio Rate
& Vio Prec.
& Leak. Rec. \\ \midrule
Recall
& 1.38
& 76.0
& 39.3 \\
\masp-Lite 
& 1.34 
& \textbf{85.4}
& 51.1 \\ 
\masp-Lite + RAG 
& 1.48
& 78.3
& \textbf{53.6} \\ \bottomrule
\end{tabular}
}%
\label{tab:ranking-masp-lite}
\vspace{-6pt}
\end{table}

\subsection{MLMs for Content Moderation}
Multimodal language models (MLMs) have demonstrated impressive capabilities in understanding images~\cite{hudson2019gqa, goyal2017making, marino2019ok} and short videos~\cite{caba2015activitynet,patraucean2024perception,li2024mvbench,Maaz2023VideoChatGPT}. These models generally comprise three modules: a vision encoder, a large language model (LLM) decoder, and an adapter that aligns image-text modalities~\cite{li2023blip,zhu2023minigpt,Maaz2023VideoChatGPT,lin2023video,liu2024llava}. 
Prior work explores MLM-based moderation models~\cite{kuaimod} 
and data generation frameworks validated on open-source VLMs~\cite{wang2025ttvlmframework}. 
These approaches focus on offline evaluation, without considering real-world deployment at production scale, while \masp is trained under our compliance regulations and operates in full-traffic production.

\section{Conclusion}
In this paper, we present \masp, a unified vision-language model that reframes video moderation from fragmented black-box classifiers to an interpretable, caption-driven framework. 
Built upon \masp, we construct a unified moderation pipeline comprising a high-throughput Risk Filter, a Moderation Actor, and a Trend Governance module. 
This \masp-based system has been deployed on our global-scale short-video platform, operating over full video traffic under strict compliance constraints. 
In our simulated production experiments, it achieves a 42.7\% relative reduction in violation leakage and a 37.0\% reduction in overkill rate, while attributing 81\% accuracy on the Advertisement downstream business.

\section*{Ethics Considerations}

The deployment of \masp for industrial-scale video moderation involves handling highly sensitive content, necessitating rigorous ethical and operational safeguards. First, we prioritize annotator welfare by enforcing strict daily exposure limits to violative content. Our Standard Operating Procedure (SOP) includes regular rotation shifts to prevent prolonged psychological strain and ensure mental well-being. For particularly sensitive categories like Child Sexual Abuse Material (CSAM), we implement dedicated protocols where access is restricted to a minimal, specially trained group of annotators operating under enhanced security and psychological support frameworks.

Regarding data privacy and retention, our system adheres to strict internal compliance standards. To minimize data footprint, deleted and private videos are permanently purged from our systems within a strictly defined retention window, while non-violative video data is retained only for a limited period. All data used for training and evaluation is stored in siloed environments with internal access controls. Any identified illegal content is handled through our internal reporting channels without permanent retention in any environment.

To prevent dual-use and adversarial risks, we have established a strict non-disclosure policy for our technical assets. We will not release the source code, model weights, or the specific training datasets used in this paper. Instead, we provide an overview of the system architecture and training objectives without offering a direct template for adversarial attacks. We are committed to ensuring that the specific enforcement logic remains protected against exploitation by bad actors.

Finally, we address bias and algorithmic fairness through our data recipe and model design. During the annotation process, our guidelines explicitly instruct annotators to avoid using or emphasizing terms related to specific ethnic groups to prevent the encoding of societal prejudices. Furthermore, we are committed to the continuous improvement of the model's multilingual capabilities, ensuring that safety enforcement is equitable and consistent across different linguistic and cultural contexts.
\section*{Limitations}

Currently, \masp does not incorporate reinforcement learning methods such as Group Relative Policy Optimization (GRPO). The policy-aware captions can serve as a form of reasoning trace indicating how a video may violate specific policies; however, platform policy guidelines are not directly encoded as reward signals and therefore are not explicitly bound to this reasoning process. 

Our system processes videos via frame sampling rather than full temporal modeling. As a result, videos that embed policy-violating content in only a single frame may evade detection if those frames are not selected. Therefore, the effectiveness of our system depends on keyframe selection quality, which remains a practical limitation under latency and computational constraints.


\clearpage

\bibliography{custom}

@inproceedings{shi2024cpfd,
  title={CPFD: Confidence-aware Privileged Feature Distillation for Short Video Classification},
  author={Shi, Jinghao and Shen, Xiang and Zhao, Kaili and Wang, Xuedong and Wen, Vera and Wang, Zixuan and Wu, Yifan and Zhang, Zhixin},
  booktitle={Proceedings of the 33rd ACM International Conference on Information and Knowledge Management},
  pages={4866--4873},
  year={2024}
}

@inproceedings{wang2025filter,
  title={Filter-And-Refine: A MLLM Based Cascade System for Industrial-Scale Video Content Moderation},
  author={Wang, Zixuan and Shi, Jinghao and Liang, Hanzhong and Shen, Xiang and Wen, Vera and Chen, Zhiqian and Wu, Yifan and Zhang, Zhixin and Xiong, Hongyu},
  booktitle={Proceedings of the 63rd Annual Meeting of the Association for Computational Linguistics (Volume 6: Industry Track)},
  pages={873--880},
  year={2025}
}

@inproceedings{liang2025embedding,
  title={Embedding-based Retrieval in Multi-Modal Content Moderation},
  author={Liang, Hanzhong and Shi, Jinghao and Shen, Xiang and Wang, Zixuan and Wen, Vera and Mehrani, Ardalan and Chen, Zhiqian and Wu, Yifan and Zhang, Zhixin},
  booktitle={Proceedings of the 48th International ACM SIGIR Conference on Research and Development in Information Retrieval},
  pages={4264--4268},
  year={2025}
}

@article{levi2025ai,
  title={AI vs. Human Moderators: A Comparative Evaluation of Multimodal LLMs in Content Moderation for Brand Safety},
  author={Adi Levi and Or Levi and Sardhendu Mishra and Jonathan Morra},
  journal={2025 IEEE/CVF International Conference on Computer Vision Workshops (ICCVW)},
  year={2025},
  pages={6024-6032},
  url={https://api.semanticscholar.org/CorpusID:280545981}
}

@article{bao2025vmoda,
  title={VModA: An Effective Framework for Adaptive NSFW Image Moderation},
  author={Bao, Han and Wang, Qinying and Chen, Zhi and Li, Qingming and Zhang, Xuhong and Li, Changjiang and Wang, Zonghui and Ji, Shouling and Chen, Wenzhi},
  journal={arXiv preprint arXiv:2505.23386},
  year={2025}
}

@article{Wang2026FromNM,
  title={From Native Memes to Global Moderation: Cros-Cultural Evaluation of Vision-Language Models for Hateful Meme Detection},
  author={Mo Wang and Kaixuan Ren and Pratik Jalan and Ahmed Ashraf and Tuong Vy Vu and Rahul Seetharaman and Shah Nawaz and Usman Naseem},
  year={2026},
  url={https://api.semanticscholar.org/CorpusID:285451876},
  journal={arXiv preprint arXiv:2602.07497}
}

@misc{huang2025llavashieldsafeguardingmultimodalmultiturn,
      title={LLaVAShield: Safeguarding Multimodal Multi-Turn Dialogues in Vision-Language Models}, 
      author={Guolei Huang and Qinzhi Peng and Gan Xu and Yuxuan Lu and Yongjun Shen},
      year={2025},
      eprint={2509.25896},
      archivePrefix={arXiv},
      primaryClass={cs.CV},
      url={https://arxiv.org/abs/2509.25896}, 
}

@article{wang2025vips,
  title={VIPS: Learning-View-Invariant Feature for Person Search},
  author={Wang, Hexu and Luo, Wenlong and Wu, Wei and Xie, Fei and Liu, Jindong and Li, Jing and Zhang, Shizhou},
  journal={Sensors},
  volume={25},
  number={17},
  pages={5362},
  year={2025},
  publisher={MDPI}
}

@article{lee2025vision,
  title={Are Vision-Language Models Safe in the Wild? A Meme-Based Benchmark Study},
  author={Lee, DongGeon and Jang, Joonwon and Jeong, Jihae and Yu, Hwanjo},
  journal={arXiv preprint arXiv:2505.15389},
  year={2025}
}

@misc{sun2023evaclipimprovedtrainingtechniques,
      title={EVA-CLIP: Improved Training Techniques for CLIP at Scale}, 
      author={Quan Sun and Yuxin Fang and Ledell Wu and Xinlong Wang and Yue Cao},
      year={2023},
      eprint={2303.15389},
      archivePrefix={arXiv},
      primaryClass={cs.CV},
      url={https://arxiv.org/abs/2303.15389}, 
}

@article{comanici2025gemini,
  title={Gemini 2.5: Pushing the frontier with advanced reasoning, multimodality, long context, and next generation agentic capabilities},
  author={Comanici, Gheorghe and Bieber, Eric and Schaekermann, Mike and Pasupat, Ice and Sachdeva, Noveen and Dhillon, Inderjit and Blistein, Marcel and Ram, Ori and Zhang, Dan and Rosen, Evan and others},
  journal={arXiv preprint arXiv:2507.06261},
  year={2025}
}

@article{Jiang2023Mistral7,
  title={Mistral 7B},
  author={Albert Qiaochu Jiang and Alexandre Sablayrolles and Arthur Mensch and Chris Bamford and Devendra Singh Chaplot and Diego de Las Casas and Florian Bressand and Gianna Lengyel and Guillaume Lample and Lucile Saulnier and L{\'e}lio Renard Lavaud and Marie-Anne Lachaux and Pierre Stock and Teven Le Scao and Thibaut Lavril and Thomas Wang and Timoth{\'e}e Lacroix and William El Sayed},
  journal={ArXiv},
  year={2023},
  volume={abs/2310.06825},
  url={https://api.semanticscholar.org/CorpusID:263830494}
}

@inproceedings{kuaimod,
  title={Vlm as policy: Common-law content moderation framework for short video platform},
  author={Lu, Xingyu and Zhang, Tianke and Meng, Chang and Wang, Xiaobei and Wang, Jinpeng and Zhang, Yi-Fan and Tang, Shisong and Liu, Changyi and Ding, Haojie and Jiang, Kaiyu and others},
  booktitle={Proceedings of the 31st ACM SIGKDD Conference on Knowledge Discovery and Data Mining V. 2},
  pages={4682--4693},
  year={2025}
}

@inproceedings{wang2025ttvlmframework,
  title={Reasoning-Enhanced Domain-Adaptive Pretraining of Multimodal Large Language Models for Short Video Content Governance},
  author={Wang, Zixuan and Sun, Yu and Wang, Hongwei and Jing, Baoyu and Shen, Xiang and Dong, Xin Luna and Hao, Zhuolin and Xiong, Hongyu and Song, Yang},
  booktitle={Proceedings of the 2025 Conference on Empirical Methods in Natural Language Processing: Industry Track},
  pages={1104--1112},
  year={2025}
}

@article{dream1k,
  title={Tarsier: Recipes for training and evaluating large video description models},
  author={Wang, Jiawei and Yuan, Liping and Zhang, Yuchen and Sun, Haomiao},
  journal={arXiv preprint arXiv:2407.00634},
  year={2024}
}

@article{Gongane2022survey,
title={Detection and moderation of detrimental content on social media platforms: current status and future directions},
author={Vaishali U. Gongane and M. Munot and A. Anuse},
journal={Social Network Analysis and Mining},
year={2022},
volume={12},
doi={10.1007/s13278-022-00951-3}
}

@article{Gorwa2020Algorithmic,title={Algorithmic content moderation: Technical and political challenges in the automation of platform governance},author={Robert Gorwa and Reuben Binns and Christian Katzenbach},journal={Big Data \& Society},year={2020},volume={7},doi={10.1177/2053951719897945}}

@article{Udupa2023Ethical,title={Ethical scaling for content moderation: Extreme speech and the (in)significance of artificial intelligence},author={Sahana Udupa and Antonis Maronikolakis and Axel Wisiorek},journal={Big Data \& Society},year={2023},volume={10},doi={10.1177/20539517231172424}}

@article{Lai2022Human-AI,title={Human-AI Collaboration via Conditional Delegation: A Case Study of Content Moderation},author={Vivian Lai and Samuel Carton and Rajat Bhatnagar and Vera Liao and Yunfeng Zhang and Chenhao Tan and Q. Liao},journal={Proceedings of the 2022 CHI Conference on Human Factors in Computing Systems},year={2022},doi={10.1145/3491102.3501999}}

@article{Yousaf2022A,
title={A Deep Learning-Based Approach for Inappropriate Content Detection and Classification of YouTube Videos},
author={Kanwal Yousaf and Tabassam Nawaz},
journal={IEEE Access},
year={2022},
volume={10},
pages={16283-16298},
doi={10.1109/access.2022.3147519}
}

@article{Naseeb2025Machine,title={Machine Learning- and Deep Learning-Based Multi-Model System for Hate Speech Detection on Facebook},author={Amna Naseeb and Muhammad Zain and Nisar Hussain and Amna Qasim and Fiaz Ahmad and Grigori Sidorov and A. Gelbukh},journal={Algorithms},year={2025},doi={10.3390/a18060331}}

@article{Wang2023Validating,
title={Validating Multimedia Content Moderation Software via Semantic Fusion},
author={Wenxuan Wang and Jingyuan Huang and Chang Chen and Jiazhen Gu and Jianping Zhang and Weibin Wu and Pinjia He and Michael R. Lyu},
journal={Proceedings of the 32nd ACM SIGSOFT International Symposium on Software Testing and Analysis},
year={2023},
doi={10.1145/3597926.3598079}
}

@inproceedings{li2023blip,
  title={Blip-2: Bootstrapping language-image pre-training with frozen image encoders and large language models},
  author={Li, Junnan and Li, Dongxu and Savarese, Silvio and Hoi, Steven},
  booktitle={International conference on machine learning},
  pages={19730--19742},
  year={2023},
  organization={PMLR}
}

@article{zhu2023minigpt,
  title={MiniGPT-4: Enhancing Vision-Language Understanding with Advanced Large Language Models},
  author={Zhu, Deyao and Chen, Jun and Shen, Xiaoqian and Li, Xiang and Elhoseiny, Mohamed},
  journal={arXiv preprint arXiv:2304.10592},
  year={2023}
}

@inproceedings{Maaz2023VideoChatGPT,
    title={Video-ChatGPT: Towards Detailed Video Understanding via Large Vision and Language Models},
    author={Maaz, Muhammad and Rasheed, Hanoona and Khan, Salman and Khan, Fahad Shahbaz},
    booktitle={Proceedings of the 62nd Annual Meeting of the Association for Computational Linguistics (ACL 2024)},
    year={2024}
}

@article{lin2023video,
  title={Video-llava: Learning united visual representation by alignment before projection},
  author={Lin, Bin and Zhu, Bin and Ye, Yang and Ning, Munan and Jin, Peng and Yuan, Li},
  journal={arXiv preprint arXiv:2311.10122},
  year={2023}
}

@inproceedings{liu2024improved,
  title={Improved baselines with visual instruction tuning},
  author={Liu, Haotian and Li, Chunyuan and Li, Yuheng and Lee, Yong Jae},
  booktitle={Proceedings of the IEEE/CVF conference on computer vision and pattern recognition},
  pages={26296--26306},
  year={2024}
}

@misc{liu2024llava,
  title={Llava-next: Improved reasoning, ocr, and world knowledge},
  author={Liu, Haotian and Li, Chunyuan and Li, Yuheng and Li, Bo and Zhang, Yuanhan and Shen, Sheng and Lee, Yong Jae},
  year={2024}
}

@article{liu2023visual,
  title={Visual instruction tuning},
  author={Liu, Haotian and Li, Chunyuan and Wu, Qingyang and Lee, Yong Jae},
  journal={Advances in neural information processing systems},
  volume={36},
  pages={34892--34916},
  year={2023}
}

@article{li2024llavaov,
  	title={LLaVA-OneVision: Easy Visual Task Transfer},
  	author={Li, Bo and Zhang, Yuanhan and Guo, Dong and Zhang, Renrui and Li, Feng and Zhang, Hao and Zhang, Kaichen and Li, Yanwei and Liu, Ziwei and Li, Chunyuan},
  	journal={arXiv preprint arXiv:2408.03326},
  	year={2024}
}

@article{gpt4o,
  author = {OpenAI},
  bibsource = {dblp computer science bibliography, https://dblp.org},
  doi = {10.48550/ARXIV.2303.08774},
  eprint = {2303.08774},
  eprinttype = {arXiv},
  journal = {CoRR},
  title = {{GPT-4} Technical Report},
  url = {https://doi.org/10.48550/arXiv.2303.08774},
  volume = {abs/2303.08774},
  year = 2023
}

@inproceedings{goyal2017making,
  title={Making the v in vqa matter: Elevating the role of image understanding in visual question answering},
  author={Goyal, Yash and Khot, Tejas and Summers-Stay, Douglas and Batra, Dhruv and Parikh, Devi},
  booktitle={Proceedings of the IEEE conference on computer vision and pattern recognition},
  pages={6904--6913},
  year={2017}
}

@inproceedings{marino2019ok,
  title={Ok-vqa: A visual question answering benchmark requiring external knowledge},
  author={Marino, Kenneth and Rastegari, Mohammad and Farhadi, Ali and Mottaghi, Roozbeh},
  booktitle={Proceedings of the IEEE/cvf conference on computer vision and pattern recognition},
  pages={3195--3204},
  year={2019}
}

@inproceedings{hudson2019gqa,
  title={Gqa: A new dataset for real-world visual reasoning and compositional question answering},
  author={Hudson, Drew A and Manning, Christopher D},
  booktitle={Proceedings of the IEEE/CVF conference on computer vision and pattern recognition},
  pages={6700--6709},
  year={2019}
}

@inproceedings{caba2015activitynet,
  title={Activitynet: A large-scale video benchmark for human activity understanding},
  author={Caba Heilbron, Fabian and Escorcia, Victor and Ghanem, Bernard and Carlos Niebles, Juan},
  booktitle={Proceedings of the ieee conference on computer vision and pattern recognition},
  pages={961--970},
  year={2015}
}

@inproceedings{li2024mvbench,
  title={Mvbench: A comprehensive multi-modal video understanding benchmark},
  author={Li, Kunchang and Wang, Yali and He, Yinan and Li, Yizhuo and Wang, Yi and Liu, Yi and Wang, Zun and Xu, Jilan and Chen, Guo and Luo, Ping and others},
  booktitle={Proceedings of the IEEE/CVF Conference on Computer Vision and Pattern Recognition},
  pages={22195--22206},
  year={2024}
}

@article{patraucean2024perception,
  title={Perception test: A diagnostic benchmark for multimodal video models},
  author={Patraucean, Viorica and Smaira, Lucas and Gupta, Ankush and Recasens, Adria and Markeeva, Larisa and Banarse, Dylan and Koppula, Skanda and Malinowski, Mateusz and Yang, Yi and Doersch, Carl and others},
  journal={Advances in Neural Information Processing Systems},
  volume={36},
  pages={42748--42761},
  year={2023}
}

@inproceedings{LLaVA-OneVision-1.5,
  title={LLaVA-OneVision-1.5: Fully Open Framework for Democratized Multimodal Training},
  author={An, Xiang and Xie, Yin and Yang, Kaicheng and Zhang, Wenkang and Zhao, Xiuwei and Cheng, Zheng and Wang, Yirui and Xu, Songcen and Chen, Changrui and Wu, Chunsheng and Tan, Huajie and Li, Chunyuan and Yang, Jing and Yu, Jie and Wang, Xiyao and Qin, Bin and Wang, Yumeng and Yan, Zizhen and Feng, Ziyong and Liu, Ziwei and Li, Bo and Deng, Jiankang},
  booktitle={arXiv},  
  year={2025}
 }

@inproceedings{sultani2018real,
  title={Real-world anomaly detection in surveillance videos},
  author={Sultani, Waqas and Chen, Chen and Shah, Mubarak},
  booktitle={Proceedings of the IEEE conference on computer vision and pattern recognition},
  pages={6479--6488},
  year={2018}
}

@inproceedings{hassner2012violent,
  title={Violent flows: Real-time detection of violent crowd behavior},
  author={Hassner, Tal and Itcher, Yossi and Kliper-Gross, Orit},
  booktitle={2012 IEEE computer society conference on computer vision and pattern recognition workshops},
  pages={1--6},
  year={2012},
  organization={IEEE}
}

@article{videosafetybench,
  title={Video-safetybench: A benchmark for safety evaluation of video lvlms},
  author={Liu, Xuannan and Li, Zekun and He, Zheqi and Li, Peipei and Xia, Shuhan and Cui, Xing and Huang, Huaibo and Yang, Xi and He, Ran},
  journal={arXiv preprint arXiv:2505.11842},
  year={2025}
}

\clearpage
\appendix

\section{\masp}

\subsection{Training Data}\label{app:training-data}

\begin{figure*}[t] 
  \centering
  \includegraphics[width=\linewidth]{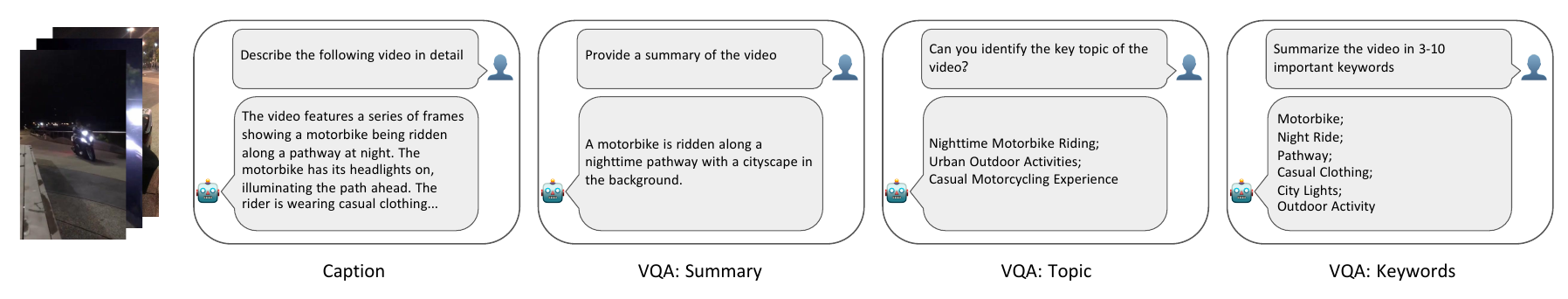}
  \caption{Examples of \masp training data. We introduce four different tasks: Caption, Summary, Topic, and Keywords. }
  \vspace{-2pt}
  \label{fig:training-data-example}
\end{figure*}

A series of training data examples are provided in Figure \ref{fig:training-data-example}.

\subsection{Training Details}
We report our training hyperparameters in Table \ref{tab:hyperparameters}.

\subsection{CapBench}\label{app:capbench-details}
\begin{table*}[tb]
\centering
\caption{Distribution of safety domains and policies in CapBench.}
\renewcommand{\arraystretch}{1.12}
\resizebox{0.75\textwidth}{!}{%
\begin{tabular}{@{}llc@{}}
\toprule
\textbf{Domain} & \textbf{Policy Category} & \textbf{\# Samples} \\ \midrule
Violence 
& Violent Behaviors; Shocking \& Graphic Content 
& 2{,}700 \\

Sexual Abuse 
& Exploitation \& Abuse; Nudity \& Sexual Activity 
& 4{,}339 \\

Mental Health 
& Mental Health; Harassment \& Hateful Behavior 
& 1{,}248 \\

Regulated Activity 
& High-Risk \& Regulated Activities 
& 4{,}656 \\

Integrity 
& Harmful Misinformation; Deceptive Behaviors 
& 624 \\ \bottomrule
\end{tabular}
}
\label{tab:capbench-stats}
\vspace{-2pt}
\end{table*}

We show the detailed statistics of CapBench in Table~\ref{tab:capbench-stats}. 
Note that since a video can violate multiple policies simultaneously, the sum of policy-specific counts exceeds the total number of unique samples.

The evaluation prompts for both stages are presented below, including event extraction and entailment judgment.

\begin{story}[Event Extraction Prompt]{
You are given a description of a video clip:
[caption]

Your tasks:\\
1. Extract at most 10 key events from the description.\\
2. From these events, identify those that may violate the video platform content policy labels: [policy\_list]. Only include events that can serve as sufficient evidence of the listed policy violations. You will be given a document of policy instructions to help you better understand the policy definition (if any).\\

Requirements for Events:\\
- An event must include an action or motion. \\
- Merge semantically similar and disjunctive actions into a single event. DO NOT repeat same events.\\
- Every event is represented by a brief sentence within 20 words, with a subject with key attributes, a predicate and optionally an object. Include the action or attributes (such as minor's age, revealing outfit) that may violate the content policy!\\
- Every event must be atomic, meaning that it cannot be further split into multiple events.\\
- Scene cuts and camera movements are NOT events!
- Substitute pronouns by the nouns they refer to.\\

Output Format:\\
Return a Python dictionary string with the following keys:\\
- ``events'': List[str] \\ the list of all extracted events.\\
- ``violative\_events'': Dict[str, List[str]] \\
mapping from each policy name in policy\_list to the list of events that violate it. If no events violate a policy, use an empty list.\\

Materials:\\
Policy Instructions:
[policy\_doc]
}
\end{story}

\begin{story}[Entailemnet Judgement Prompt]{
You are given a video description and a list of events.

Your task is to classify the relationship between the video description and each event into one of three classes:
- entailment: The video description entails the event.\\
- contradiction: Some detail in the video description contradicts the event.\\
- neutral: The relationship is neither entailment nor contradiction.\\

Video Description: [caption]\\
Events:[events]\\

Output Format:\\
Return a JSON list in the following format:\\
\{"event": "copy an event here", \\
  "relationship": "put class name here"\}\\

Requirements:
- Classify every event independently.\\
- Use only one of the three predefined class names.\\
- Do NOT provide any additional text, explanation, or formatting.\\
- Output only the JSON.
}
\end{story}

\begin{figure}[t] 
  \centering
  \includegraphics[width=\linewidth]{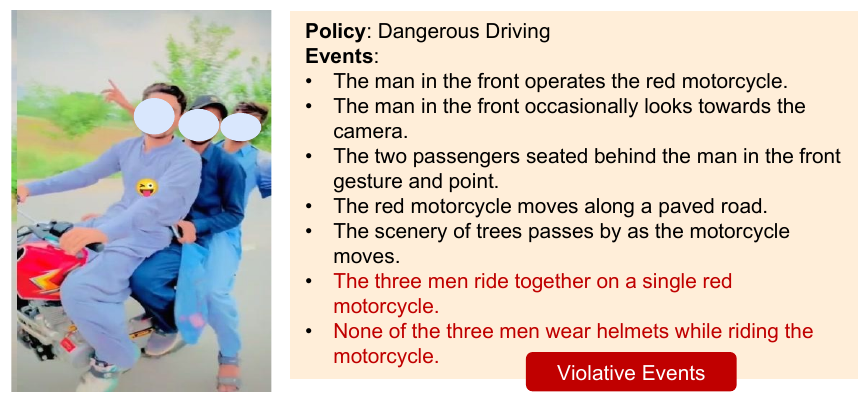}
  \caption{CapBench example.}
  \vspace{-2pt}
  \label{fig:capbench-example}
\end{figure}

\subsection{Ablation Study}\label{app:masp-model-ablation}

As shown in Table~\ref{tab:masp-ablation}, we also examine the impact of visual and resolution components. Intuitively, replacing the vision encoder with less performant EvaCLIP~\cite{sun2023evaclipimprovedtrainingtechniques} results in a decrease in violation recall (-9.5\%). 
For resolution setting, given the on-par performance but substantially higher token cost, we choose not to use AnyRes in our model design.

\section{\masp-Lite}

As shown in Appendix~\ref{app:masplite-prompt}, we fuse video frames with auxiliary textual features as model input, including titles, user profiles, OCR, and ASR, to provide enriched semantic context. 

\subsection{Case Study}\label{app:masplite-casestudy}
We find that \masp-Lite exhibits a robust capacity for contextual inference, enabling the detection of latent risk signals that extend beyond explicit keywords. In High-Risk Driving scenarios, the model identifies danger by synthesizing environmental cues (such as dashboard speedometers) even in the absence of overt reckless actions. Similarly, in Severe Bullying cases, \masp-Lite effectively perceives hostile intent conveyed through suggestive or indirect language rather than explicit insults. Furthermore, in the domain of Serious Harm, the model recognizes subtle precursors to danger, successfully identifying implicit threats that often precede manifest harmful behavior.

\section{\masp-RAG}
We show the example violative events in Table~\ref{tab:violative-event-examples}.

\begin{table*}[tb]
\centering
\caption{Examples of violative events from the knowledge base for \masp-RAG.}
\renewcommand{\arraystretch}{1.25}
\begin{tabular}{@{}p{0.32\textwidth} p{0.65\textwidth}@{}}
\toprule
\textbf{Policy} & \textbf{Violative Events} \\ \midrule

Nudity \& Sexual Activity 
& The video shows two youths performing a dance that includes sexually suggestive movements, such as hip thrusting, squatting, and sticking their tongues out. \\

High-Risk \& Regulated Activities 
& The video shows multiple people riding on the exterior of moving vehicles, including on the sides and tops of trucks within a convoy. \\

Mental Health 
& The video contains a real-life clip of a person standing on a train platform as a train passes, followed by animated scenes depicting bloodstains on the platform. \\

Harassment \& Hateful Behavior 
& The video displays a person's full email address (xxx@gmail.com) and a verification code on a smartphone screen. \\

\bottomrule
\end{tabular}
\label{tab:violative-event-examples}
\vspace{-2pt}
\end{table*}

\begin{table*}[t]
\centering
\caption{Ablation results of \masp-7B on CapBench.}
\vspace{-3pt}
\renewcommand{\arraystretch}{1.12}
\resizebox{\textwidth}{!}{
\begin{tabular}{lccccccccc}
\toprule
& \multicolumn{5}{c}{Violative Set} 
& \multicolumn{4}{c}{Healthy Set} \\ 

Model 
& Violence 
& Sex Abuse 
& Mental Health 
& Regulated Act 
& Integrity 
& Non-vio Rec. 
& Rec. 
& Prec. 
& F1 \\ 
\cmidrule(lr){2-6} \cmidrule(lr){7-10}

\masp-7B        
& 56.3 & 51.3 & 50.2 & 57.7 & 50.1 
& 32.4 
& 28.9 & 82.3 & 39.1 \\

w/ EvaCLIP      
& 46.6 & 44.2 & 39.4 & 46.9 & 35.8 
& 24.3 
& 21.6 & 77.1 & 30.0 \\

w/ AnyRes       
& 56.7 & 51.4 & 48.9 & 58.4 & 50.6 
& 32.5 
& 28.7 & 82.4 & 38.7 \\

\bottomrule
\end{tabular}
}
\vspace{-6pt}
\label{tab:masp-ablation}
\end{table*}

\begin{table}[tb]
\centering
\caption{\masp training hyper-parameters}
\label{tab:hyperparameters}
\begin{tabular}{lc}
\toprule
\textbf{Hyperparameter} & \textbf{Value} \\ 
\midrule
\# Epochs        & 2              \\
Per-device Batch Size   & 8              \\
Gradient Accumulation   & 2              \\
Learning Rate           & 1e-5           \\
LR Scheduler            & Cosine         \\
Warmup Ratio            & 0.03           \\
\bottomrule
\vspace{-5pt}
\end{tabular}
\end{table}

\section{Prompts}
\label{sec:appendix-prompts}

\subsection{Evaluation}

The captioner prompt for Capbench is as follows:

\begin{story}[User Prompt]{
Given image frames uniformly sampled from a video clip, describe the video (not the individual images) in detail, focusing on the main subjects, their actions, and the background scene. You should also pay attention to details that may violate the trust-and-safety platform content policy. Don't describe feelings or atmosphere.

Your output should be a single coherent paragraph. Maximum length: 150 words.
}
\end{story}

\subsection{\masp-Lite Inference} \label{app:masplite-prompt}

The online inference prompt of \masplite is as follows:

\begin{story}[User Prompt]{
Region: [region]\\
Title: [title]\\
User nickname: [nickname]\\
ASR Text: [asr]\\
OCR Text: [ocr]\\
Bio Text: [profie]\\

Based on the video frames and related content above, please indicate the severity of any inappropriate, disruptive, or harmful content.
}
\end{story}

\subsection{\masp-RAG Inference}\label{app:masprag-prompt}

The online inference prompt of \masp-RAG is as follows:

\begin{story}[User Prompt]{
Region: [region]\\
Title: [title]\\
User nickname: [nickname]\\
ASR Text: [asr]\\
OCR Text: [ocr]\\
Bio Text: [profie]\\

Violative Events for Reference:\\
1. [policy title], violation reason: [reason]\\
2. [policy title], violation reason: [reason]\\
3. [policy title], violation reason: [reason]\\
    
Based on the video frames, the above content, and the judgment logic of the similar cases, please indicate the severity of any inappropriate, disruptive, or harmful content.
}
\end{story}

\end{document}